**Searching for extraterrestrial intelligence signals in astronomical spectra, including existing data**

Ermanno F. Borra,

Centre d'Optique, Photonique et Laser,

Département de Physique, Université Laval, Québec, Qc, Canada G1K 7P4

(email: borra@phy.ulaval.ca)

**SHORT TITLE:** Searching for ETI in spectra

RECEIVED________________________________

**ABSTRACT**

The main purpose of this article is to make Astronomers aware that Searches for Extraterrestrial Intelligence can be carried out by analyzing standard astronomical spectra, including those they already have taken. Simplicity is the outstanding advantage of a search in spectra. The spectra can be analyzed by simple eye inspection or a few lines of code that uses Fourier transform software. Theory, confirmed by published experiments, shows that periodic signals in spectra can be easily generated by sending light pulses separated by constant time intervals. While part of this article, like all articles on searches for ETI, is highly speculative the basic physics is sound. In particular, technology now available on Earth could be used to send signals having the required energy to be detected at a target located 1000 light years away. Extraterrestrial Intelligence (ETI) could use these signals to make us aware of their existence. For an ETI, the technique would also have the advantage that the signals could be detected both in spectra and searches for intensity pulses like those currently carried out on Earth.

# 1. INTRODUCTION

The Search for Extraterrestrial Intelligence (SETI) is a recent scientific endeavor. The first published suggestion was made by Cocconi & Morrison (1959) and interest has gradually increased within the scientific community. There presently are several dedicated SETI projects now underway. An article in Annual Reviews of Astronomy & Astrophysics by Tarter (2001) gives a summary of the history and the relevant physical and sociological issues in SETI. The review concludes that there presently is no evidence whatsoever for or against the existence of ETI. However, as Tarter (2001) mentions in the introduction, SETI is a long shot endeavor with an immense payoff: Arguably finding evidence of an ETI would be one of the most important event in the history of humanity.

The first astronomical searches for ETI were carried out with radio telescopes. Following the suggestion by Schwartz & Townes (1961) and Townes (1983) that infrared and optical lasers could be used for interstellar communications, SETI has begun in the optical region. For example, the 1.5-m diameter Wyeth Telescope at the Harvard/Smithsonian Oak Ridge Observatory carries out searches for nanosecond pulses from ETI (Howard et al. 2004). Howard et al. (2004) discuss in details the issues relevant to a search for nanosecond pulses. A recent article by Korpela et al. (2011) summarizes the status of the UC-Berkeley SETI effort which includes radio and optical telescopes. The SEVENDIP instrument at UC-Berkeley uses an automated 0.8-meter telescope to search for nanosecond pulses in the 300-700 nm wavelength region (Korpela et al 2011).

There also have been suggestions of searches for ETI in astronomical spectra (e.g. by Whitmire & Wright (1980) and by Paprotny (1977)). They suggested searching for anomalous spectral lines originating from radioactive fissile waste material. Reines & Marcy (2002) searched, in 577 nearby stars, for emission lines too narrow to natural from the host star, like lines originating from lasers.

Present techniques used in optical SETI to measure intensity time variations have several limitations. They can only observe one object at a time. They are limited to bright objects. Their major inconvenience is that they need dedicated instruments or require precious telescope time on standard telescopes.

Borra (2010) shows that periodic time variations of the intensity signal originating from a pulsating source modulate its frequency spectrum with periodic structures. Periodic time variations of the intensity signal originating from a pulsating source with periods between *$10^{-10}$* and *$10^{-15}$* seconds would modulate its spectrum with periodic structures detectable in standard astronomical spectra. Periods shorter than *$10^{-10}$* seconds could be detected in high-resolution spectra. Note that the modulation is rigorously periodic in the frequency units spectrum but not in the wavelength units spectrum.

In this article I suggest that searches for extraterrestrial intelligence should be carried out by analyzing astronomical spectra, including spectra already taken. The outstanding advantage of the technique is that it does not require any specialized equipment: To the contrary, one can use existing spectroscopic data acquired for other purposes. The data analysis technique is also extremely simple. The data can be analyzed by direct eye inspection or with simple Fourier transform software.

## 2. SIGNALS FROM EXTRA-TERRESTRIAL INTELLIGENCE

While some of the following discussion, like all SETI, is highly speculative, it is based on proven Physics. Consider an extraterrestrial civilization which is more advanced than ours. They have decided that they want to signal their existence to other civilizations. This is not an easy task considering how vast the universe is. The best chance of being seen is by being accidentally detected during astronomical observations. They know, on the basis of their own experience, that a scientifically advanced civilization observes the sky and takes spectra of astronomical objects. In particular, spectroscopic surveys are carried out. Tarter (2001) makes similar statements and emphasizes the importance of using surveys. Consequently, a good way to let others know of their existence is to generate a signal that is so unusual that it can only be artificial (Tarter 2001). A most unusual signal would be made of a spectral modulation of the spectrum that is so unusual that it warrants more observations which will then reveal that it is artificial.

A spectral modulation can readily be generated by sending short bursts of light (e.g. with a powerful laser). Borra (2010) uses Fourier transform theory to show that a pulsating source sending a time dependent electric field $E(t)$ made of $N$ pulses separated by the time $\tau$ generates a spectral distribution given by

$$S(\nu) = S_1(\nu)\left[\sin(2\pi\nu N\tau/2)/(\sin(2\pi\nu\tau/2))\right]^2, \qquad (1)$$

where $\nu$ is the frequency of observation and $S_1(\nu)$ is the continuous spectrum of a single pulse, which therefore depends on the duration of the pulse. For large values of $N$ Equation (1) predicts spectral shapes having sharp peaks separated *by $1/\tau$*, resembling comb functions made of $\delta$ functions. Note that Borra (2010) uses angular frequencies $\omega = 2\pi\nu$.

The theoretical analysis leading to Equation (1) is experimentally supported by the experiments of Chin et al. (1992). They used a grating spectrometer to measure the spectral modulation in the visible wavelength region caused by pairs of 150 femtosecond pulses separated by times $\tau$ varying between *3 $10^{-13}$* seconds and *5 $10^{-11}$* seconds. The pairs of pulses were emitted at intervals of *1.3 $10^{-8}$* seconds. The experiments of Chin et al. (1992) show the periodic modulation of $S(\omega)$ predicted by Equation (1) for 2 pulses ($N=2$)

$$S(\nu) = S_1(\nu)\left[2\cos(2\pi\nu\tau/2)\right]^2. \qquad (2)$$

Equations (1) and (2) show that the periodic frequency spacing is inversely proportional to $\tau$. Standard spectroscopic equipment in the visual-infrared regions of the spectrum would allow us to detect the spectral signatures of pulses with time separations shorter than $10^{-9}$ seconds, depending on the resolution of the spectrometer. Note however that the signal, which is periodic in frequency units is not periodic in wavelength units, as discussed in section 4 and shown in figures 1 to 4.

Clearly such a signal in an astronomical source is strange. If detected in a spectrum, the object would obviously be observed again. The ETI obviously also know this. They may thus make it look even more peculiar by inducing unnatural time variations generated by changing the time between pulses ($\tau$ in Equations 1 and 2) in an artificial sequence. There would then be little doubt that it is an ETI signal.

Equation (2) describes the modulation given by a single pair of pulses. Pairs of pulses separated by a short time $\tau_0$ could be generated periodically with a second period far greater than $\tau_0$. One could, for example, send couples of bursts separated by $\tau_0$ every $10^{-6}$ seconds. An advantage of sending powerful tight pairs of pulses separated by long time intervals is that this helps storing energy prior to sending signals where the energy is tightly concentrated in a very short time interval. The pairs of pulses separated by the short time $\tau_0$ could also be sent with several different periods $\tau_i$ greater than $\tau_0$. The advantage of this is that the signals could also be detected in SETIs that look for pulses within short time scales greater than $\tau_0$; for example Howard et al (2004) search for nanosecond pulses in the optical region.

### 3. PHYSICAL CONSIDERATIONS

The hypothetical civilizations trying to contact us are likely to be far more advanced technologically than we presently are, possibly by tens of thousands of years (Tarter 2001), so that it is difficult to predict the technology that they may use to generate the signals described by Equations (1) and (2). It is however necessary to ascertain that the needed technology does not violate basic physical laws. A major issue comes from the fact that the energy needed should not be absurdly large. We shall therefore compare the energy requirements with the technology presently available on Earth. For this, we shall use the analysis in Howard et al, (2004) who considered the energy requirements for an ETI trying to communicate with nanosecond optical pulses. They considered the feasibility of interstellar communications with technology available at the time the paper was written. They assumed communications within a 1000 light years diameter region surrounding the Earth that would contain about 1 million sun-like stars. They assumed that a diode pumped laser similar to the Helios laser designed at Lawrence Livermore National Laboratory for inertial confinement fusion would be used. They assumed 10-meter diameter telescopes to send and receive the signals. Considering a beam that would exit the transmitting telescope with a 20 mas angle, giving a 6 AU wide beam at 1000 ly, they compute that a 3 ns pulse generated by the Helios laser, would give at the receiving 10-meter telescope 1500 photons for each 3 ns pulse. Considering that the Helios laser generates pulses with a 10 Hz frequency, we see that the receiver would observe 15,000 photons/second.

Let us first consider the case where the signal is isolated in space. We shall consider in the next section and in section 5 the case where the signal generated by the Helios laser is superposed on the spectrum of the home star. Let us convert the 15,000 photons/second counts in magnitudes to estimate the strength of the signal in the domain of astronomical observations. Considering that a 10-m telescope would receive $3 \; 10^7$ photons/second from a $m_v = 12$ star (Howard et al. 2004), which is the apparent magnitude of the sun at 1000 pc, we see that the signal from the Helios laser would have a visual magnitude $m_v = 20.3$. A magnitude $m_v = 20.3$ is encouraging since a signal emitted at a distance of 1000 light years would be detectable with current technology. This signal would be detectable in a spectroscopic survey. This limiting magnitude is within reach of spectroscopic astronomical surveys like the Sloan Digital Sky Survey. Spectroscopic astronomical surveys usually target objects that are detected in imaging survey. The $m_v = 20.3$ signal would be detectable in an imaging survey. The signal would however have to be present for a sufficiently long time (e.g. several years) to be detected in both an imaging survey and a spectroscopic survey. The signal could also be accidentally detected by a slit spectrograph if the slit happens to be in the direction of the source.

It is possible that the ETI will send the signal from a location sufficiently distant from the home star that the signal would not be superposed to the spectrum of their home-star observed 1000 light years away. Note that the optical source used (e.g. a laser) does not have to be present at this location. It could be located on the home-planet and send its light beam to a space station that contains mirrors to redirect it to the targets. Assuming a 1000 ly distance (307 parsecs), a signal located at 2 arc seconds from the star, and thus detectable as a separate spectrum, would have to be located at 614 AU from the central star. This is about 20 times the semi-major axis of Pluto. It is therefore not an overwhelmingly large distance. Wiltze (2011) gives a review of the solar sails technology, estimating that a spacecraft powered by a solar sail could reach Pluto in 5 years. For an ETI at a 100 *ly* distance from Earth, the 2 arcsecond location would thus be reached within the reasonable time of 10 years. At 1000 *ly* one would need 100 years. This is a long time but not overwhelmingly so, considering that the ETI civilization may be thousands of years older than ours.

We do not know what technology ETI may use. We consider solar sails to move it to a sufficiently distant location simply to show that it could be done with today's technology and that, therefore, the suggestion is not absurd. The ETI may obviously also use better technology than solar sails.

Instead of an isolated signal, the ETI could send signal from the home planet that would therefore be superposed on the stellar spectrum. The detection of such a signal is discussed in the next section.

## 4. DETECTION OF THE SIGNAL IN ASTRONOMICAL SPECTRA, INCLUDING THE CASE WHERE THE ETI SIGNAL IS SUPERPOSED ON A STELLAR SPECTRUM

Equations 1 and 2 give the spectrum in frequency units but astronomical spectra in the infrared-optical window are usually obtained in wavelength units. While the peaks are periodic and of equal intensity in the frequency domain, they are not periodic in the wavelength domain since $\lambda = c/\nu$. Also, the peak intensity, which is constant in frequency units following Equation 1, increases in wavelength with decreasing wavelength by a $1/\lambda^2$ factor since the frequency and wavelength intervals $\Delta\nu$ and $\Delta\lambda$ are related by $\Delta\nu = c\Delta\lambda/\lambda^2$, with $\Delta\nu$ = constant.

Figure 1 in Borra (2010) shows the signals generated by Equation 1 in frequency units normalized to the same peak intensity. Figures 1 and 2 below give examples of signals generated by Equation 1 but in wavelength units. To convert from frequency to wavelength units, the intensity is increased with decreasing wavelength by a $1/\lambda^2$ factor (see previous paragraph). We assume a wavelength window similar to the window of the Sloan Digital Sky Survey (SDSS). Figure 1 illustrates the effect of changing the number of pulses *N*, while keeping the time between pulses $\tau$ constant ($\tau = 10^{-15}$ seconds). Figure 2 illustrates the effect of changing the time between pulses $\tau$ while keeping the number of pulses N ( *N*= 2 ) constant. These two figures illustrate the basic appearances of the signals in astronomical spectra and the dependence of the appearance of the signals on $N$ and $\tau$. Note in particular that the spectral features in the spectrum at the middle of Figure 1 resemble the broad emission lines of a quasar, while the spectral features in the spectrum at the bottom of Figure 1 resemble the narrow emission lines of an emission-line galaxy. If seen in a faint noisy spectrum, the spectra could be respectively misinterpreted as the spectrum of a quasar and of an emission-line galaxy. Comparing the three spectra in Figure 1, which are generated with pulses separated by the same value of $\tau = 10^{-15}$ seconds but, respectively, values $N = 2$, $N = 6$ and $N = 60$, shows the effect of the number of pulses on the shapes of the teeth of the spectral comb: The peak intensity increases and the width of the pulses decreases with increasing $N$. Note that in these figures the peak actually increases with $N^2$, but this is because the total emitted energy contained in $N$ pulses obviously increases with $N$, therefore giving another multiplication by $N$. The derivation of Equation 1 in Borra (2010)) shows that the total energy released is proportional to the number $N$ of pulses having equal energy. Consequently, to see the effect of $N$ only on the shape of the comb one must divide the arbitrary units in the Figures by $N$. Therefore, for the same total emitted energy, the peaks in the bottom spectrum of Figure 1 would have a height 10 times greater than those in the middle of Figure 1.

Equation 1 shows that pulses separated by the time $\tau$ modulate a continuous $S_1(v)$ spectrum which is given by the Fourier transform of the time dependence of the signal $V(t)$ of the individual pulses. Consequently, the wavelength interval over which the modulation of the pulses is present, and will be seen in a spectrum, depends on the duration of the pulses: the shorter the duration of the pulses, the longer the wavelength interval. For example the pulses used by Chin et al (1992) have a duration of 150 *fs* that gives a width of 4 *nm*. Since the width varies as the inverse of the duration of $S_1(v)$, durations of the order of a few *fs* would be needed to give a clearly visible modulation in a SDSS spectrum. Note that $S_1(v)$ is centered on the central wavelength of the source, as can be seen in Chin et al. (1992).

Figure 2 shows that the distance between spectral maxima increases with decreasing $\tau$. Consequently, the visibility of the spectral modulation depends on $\tau$ and the resolution of the spectrograph. Assuming a spectrograph having a resolution of 0.1 *nm*, comparable to the resolution of the SDSS, one could observe pulses separated by $10^{-12}$ s $< \tau < 10^{-15}$ s. High resolution spectrographs would allow detection of pulses separated by times as long as $\tau = 10^{-10}$ seconds Because the modulation is not periodic in the wavelength units spectrum, for a given τ, the peaks become closer with decreasing wavelength. Consequently, for large values of $\tau$ and assuming a constant wavelength resolution, the closely spaced peaks are easier to detect at longer wavelengths.

Figures 1 and 2 show the signal alone. If the signal is superposed on a stellar spectrum, the appearance of the sums of the two spectra may be less peculiar than the signal alone. Figure 3 and 4 give examples of superpositions that illustrate this. In Figure 3, the signal from a source with $N=2$ and $\tau = 5\ 10^{-15}$ seconds is superposed on the spectrum of a F9 star taken from the SDSS survey. In this particular case, a combination of noise and spectral features makes the signal difficult to detect below 5000 angstroms but make it clearly visible for wavelengths longer than 6000 angstroms. In Figure 4, the signal from a source with $N=8$ and $t = 5\ 10^{-15}$ seconds is superposed on the spectrum of the same F9 star taken from the SDSS survey. The total energy of the signal is 5% of the energy in the stellar spectrum in both cases. Note that an F9 star has very nearly the spectrum and the total energy of the sun (G2 spectrum). Because a larger number of pulses concentrate more of the energy in narrow spectral bands, the signal is much easier to detect in Figure 4 by visual inspection. On the other hand, it is easier to generate a signal made of only two pulses. Note that the total energy of the signal is the same in figures 3 and 4; while in Figure 1 the signal are generated by pulses of equal strength, so that the total signal increases with the number of pulses. For the same emitted energy, the signal would be easier to detect in a later main-sequence spectral type since a later spectral type would be considerably less luminous than an F9 star. For example, an M9 star has luminosity $10^3$ times smaller than an F9 star.

If the signal is sufficiently strong, it could be detected by simple eye inspection. For a weaker signal, and for the automated analysis of a very large number of spectra, the best way to detect a weak signal is to perform a Fourier transform in the frequency spectrum. A Fourier transform can readily extracts the periodic signal in the case where the periodic modulation is hidden in noise. Figure 5 shows a computer simulation of a flat noisy frequency spectrum with an added signal generated from Equation 2 (Equation 1 for $N = 2$ ) and $\tau = 10^{-14}$ seconds. The signal, which should contain 11 maxima and 9 minima is overwhelmed by noise and would not be detected by a simple visual inspection. Figure 6 shows the Fourier transform of the same spectrum. The signature of the periodic signal, a spike, is now clearly visible at the left of the figure. For N >2, the signal of the Fourier transform would be similar but with several spikes separated by equal values of $\tau$.

A full discussion of the parameter space that could be used to generate a signal is complex since it would involve too many parameters and is beyond the scope of this article. Figures 1 and 2 may however give us some intuitive suggestions.

## 5. DISCUSSION

Section 3 discusses the power requirements for the simple case where the spectrum is isolated in space, concluding that, assuming an Helios laser at a distance of 1000 light years from Earth, the signal, has an apparent visual magnitude $m_v = 20.3$ that could be detected in a spectroscopic astronomical survey (e.g. the SDSS survey). However, the signal could also be superposed on the spectrum of the home star. The total energy of the spectroscopic signal would then be considerably weaker (a factor of $2\ 10^3$ assuming an Helios laser) than the total energy of the stellar spectrum of a solar-type star, making it harder to detect. In a spectrum the detection could however be facilitated by increasing the contrast by appropriately choosing $N$ and $\tau$ in Equation 1. The signal predicted by $N >> 2$ is sharply peaked with the sharpness increasing with $N$ (Figure 1 shows $N = 6$ and $N = 60$) concentrating the energy in narrow spectral regions so that the emission-line-like spectral features would be easier to detect. A detailed discussion of this is complex since it depends on many parameters (e.g. $\tau$ and $N$) and is beyond the scope of this article. As illustrated by Figures 1 and 2 the half-width of the peaks not only decreases with increasing $N$, but also, for a given $N$, decreases with decreasing $\tau$.

Let us assume, for a simple short discussion, the case where $\tau = 10^{-15}$ s . Using $N = 60$ (bottom of Figure 1) would give a signal having a half-width of the order of 2 nm for 500 < λ < 600 nm. The spectral contrast would then be 2,5 % . Increasing $N$ to 600, would give a half width of 0.2 nm and a contrast of 25%, making the signal observable even in a noisy spectrum. The SDSS web site states that the spectra at g magnitudes = 20.2, which roughly corresponds to $m_v = 20$, have a signal to noise ratio between 4 and 7 per pixel (the pixels are about 0,125 nm wide). We therefore see that the 25% contrast given by the Helios laser could be detected when superposed on $m_v = 18$ spectra from the SDSS survey. Furthermore, as argued in section 3, and further discussed below, it is legitimate to assume that the hypothetical ETI would be far more technologically advanced then we presently are; consequently it is legitimate to assume that they would have considerably more powerful sources, thereby increasing the contrast and the distance at which they could be detected.

The discussion in section 3 assumes laser technology available before 2003 (Howard et al. 2004). However, as Howard et al. 2004) also suggest, we can assume that, considering the Moore's law of laser technology, a more advanced civilization should have no trouble increasing the laser power by 2 to 3 orders magnitude making the signal readily detectable. For a solar-type star at 1000 ly the signal would then be comparable to the stellar background and thus easily detectable. For an isolated beacon (e.g at 2 arcseconds from the star), the signal would have a magnitude between $15 < m_v < 13$. The Moore law suggestion is intuitively justified by simply imagining how Howard et al. (2004) and the present article would have been received before the invention of the laser 60 years ago, when the signal would have had to be generated with light bulbs!

The discussion of the energy requirements can also be carried out in terms of Kardashev ET classification (Tarter 2001). Figures 4 and 5 give examples of signals generated from Equation 1 that are superposed on an F9 stellar spectrum taken from the SDSS survey. The ET signals that are clearly visible in Figures 4 and 5 have a total observed energy that is 5% of the observed energy in the stellar spectrum. Assuming as in Howard et al. (2004) that the signal is sent with a 10-m diameter telescopes and has therefore a 20 mas angular width, we find that the energy needed for the ET signal is $10^{-16}$ times the total energy of the home star. By comparison, a Type I civilization on the Kardashev scale is capable of harnessing $10^{-10}$ times the energy of its home star. It would thus be able to generate the necessary energy but, in practice, would obviously only send a very small fraction of its total harnessed energy to a few stars. Humans presently are near a type I status. A type II civilization (humans should reach it in a few thousand years) should be capable of harnessing the energy comparable to the energy of its own star so that it would be able to easily send signals to a very large number of stars, A type III civilization (humans should reach it in about 100,000 to a million years) should harness an energy comparable to the energy of the entire galaxy and therefore be able to send very strong signals to an extremely large number of stars. Tarter (2001) suggests that civilizations trying to contact us are likely to be more advanced technologically than we presently are by tens of thousands of years and therefore should be Type II civilizations. They would therefore easily generate signals comparable to those seen in Figures 4 and 5.

## 6. CONCLUSION

ExtraTerrestrial Intelligence (ETI) could signal its existence to others by sending light pulses with time separations of the order of $10^{-9}$ to $10^{-15}$ seconds that could be detected in spectra. Signals with time separations considerably larger than nanoseconds would however be difficult to detect because the resolution of the spectroscopic equipment would be insufficient to resolve the spectroscopic signature. One also could detect spectroscopic signals from ETIs that send bursts with periodic time signals (e.g. pairs of pulses) separated by longer time scales (e.g milliseconds). The other advantage of this procedure is that the signals could also be detected in SETIs that look for intensity pulses within short time scales. For example searches for nanosecond pulses in the optical region (Howard et al (2004).

As shown in section 4, the physical requirements (e.g. energy) needed to communicate within a 1000 *ly* radius are reasonable. They could be met with lasers and telescopes presently available on Earth.

The outstanding advantage of the proposition is the simplicity of the data analysis. A strong signal could be found by visual inspection. One could also incorporate signal-finding algorithms into existing software and use it with existing databases and future spectroscopic data. As discussed in section 4, one could use Fourier transforms to detect the signal. Using a Fourier transform would be particularly useful for an automatized analysis of a large quantity of spectra (e.g. from a survey). This can be done with a few lines of code in Matlab. It is a very small effort worth doing because finding ETI, would be of enormous interest. It is a small effort with a potentially huge pay-off.

Finally the proposition meets the criterium (Tarter 2001) that new instrument that opens up pristine cell of observational phase space may surprise us with unexpected manifestations of ETI

Note also, that, although this article considers the optical-infrared spectral window, the signals could also be generated in other spectral regions (e.g. the radio region).

**ACKNOWLEDGEMENTS**

This research has been supported by the Natural Sciences and Engineering Research Council of Canada.

Whitmire, D. P.; & Wright, D. P.1980,. Icarus 42, 149

Wiltze, A. 2011, Science News 180, 18

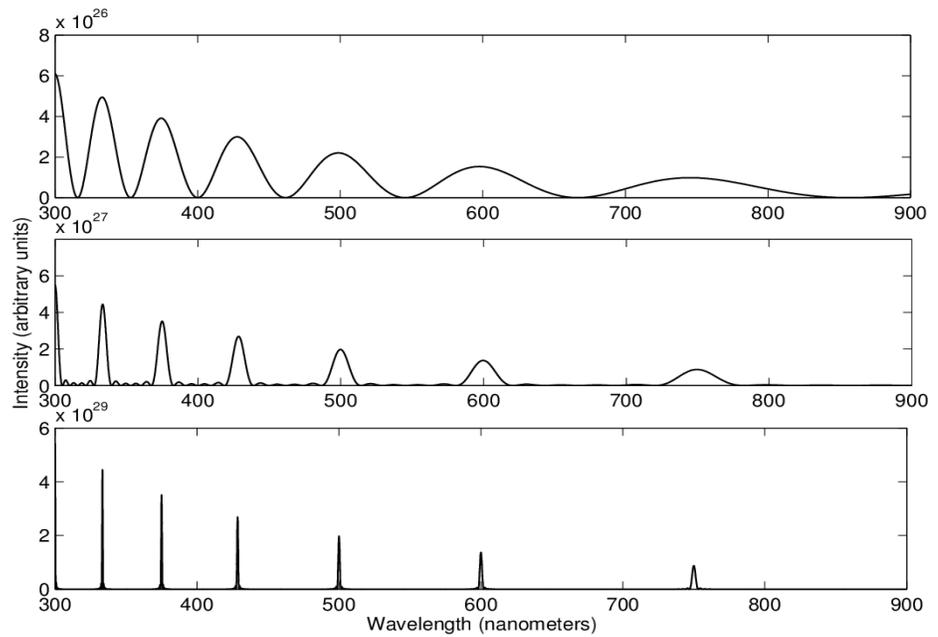

Figure 1

It illustrates the effect shows the effect, on the shapes of the teeth of the spectral comb, of changing the number of pulses $N$, while keeping the time between pulses $\tau$ constant. The three spectra are generated with pulses separated by the same value of $\tau = 10^{-15}$ *seconds but values N =2 (top spectrum), N = 6 (middle spectrum), and N = 60 (bottom spectrum).* Assuming a spectrograph having a resolution of 0.1 *nm*, comparable to the resolution of the SDSS the features would be easily resolved

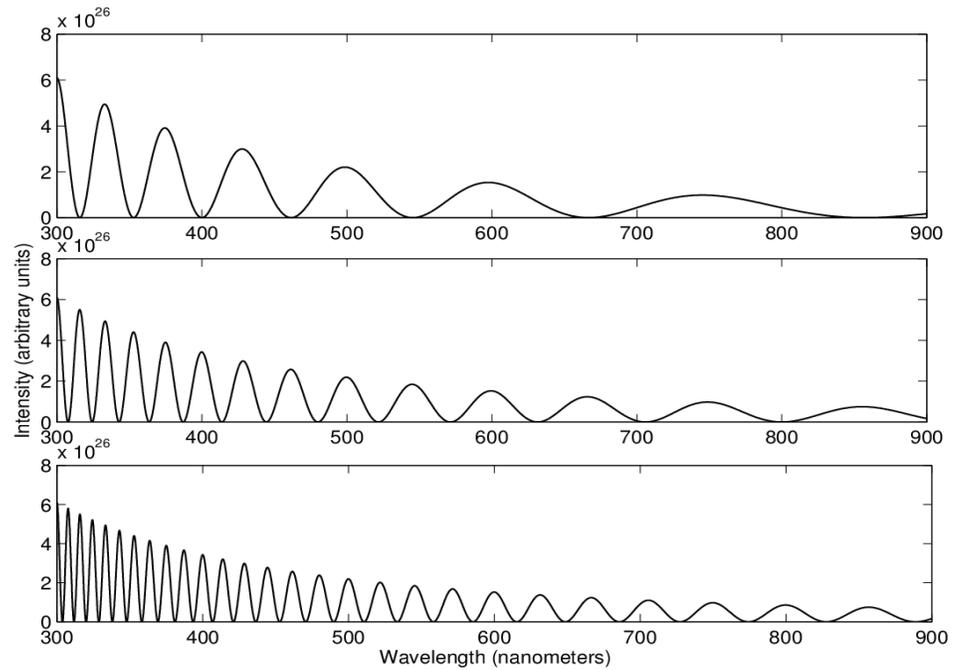

Figure 2

It illustrates the effect of changing the time between pulses $\tau$ while keeping the number of pulses $N$ constant ( $N= 2$ ). The spectrum at the top has $\tau = 10^{-15}$ seconds . the spectrum at the top has $\tau = 210^{-15}$ seconds and the spectrum at the bottom has $\tau = 410^{-15}$ seconds . Assuming a spectrograph having a resolution of 0.1 *nm*, comparable to the resolution of the SDSS the features would be easily resolved

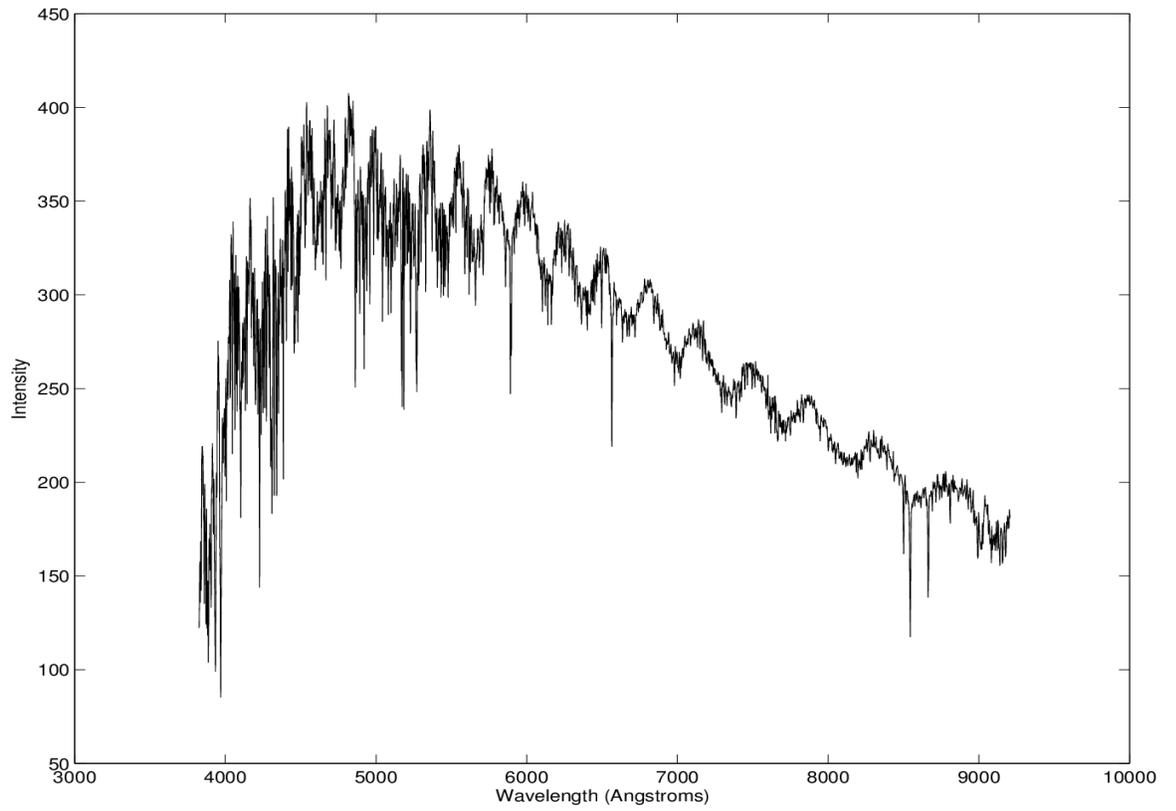

Figure 3

The signal from a source with *N=2* and $\tau = 5 \times 10^{-15}$ seconds is superposed on the spectrum of a F9 star taken from the SDSS survey,

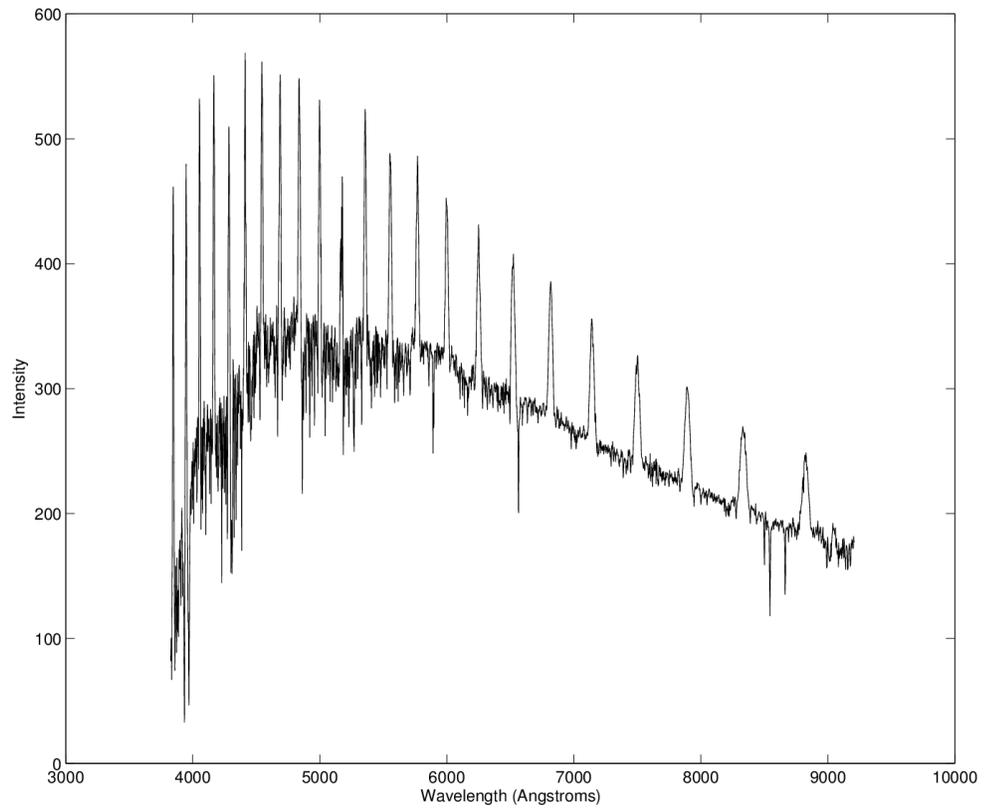

Figure 4

The signal from a source with *N=8* and *τ = 5 10$^{-15}$* seconds is superposed on the spectrum of a F9 star taken from the SDSS survey,

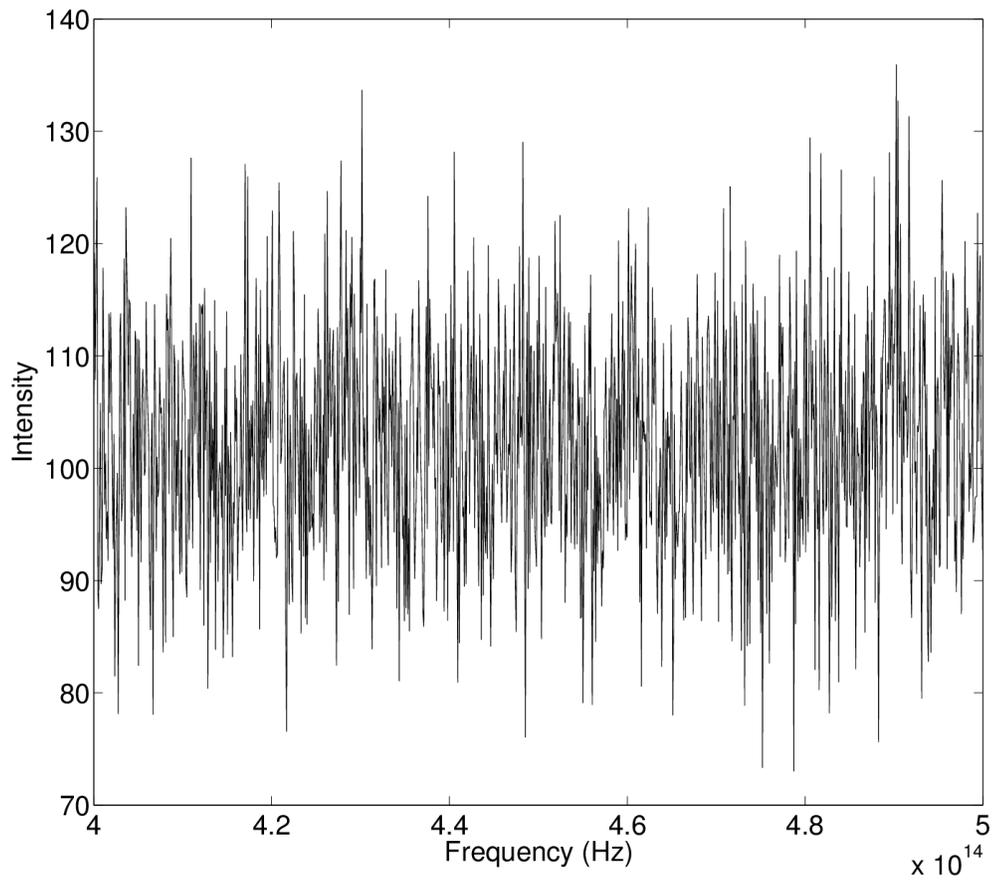

Figure 5

It shows a computer simulation of a flat noisy frequency spectrum with an added signal generated from Equation 2 for $\tau = 10^{-14}$ seconds. The signal is overwhelmed by noise and would not be detected by visual inspection.

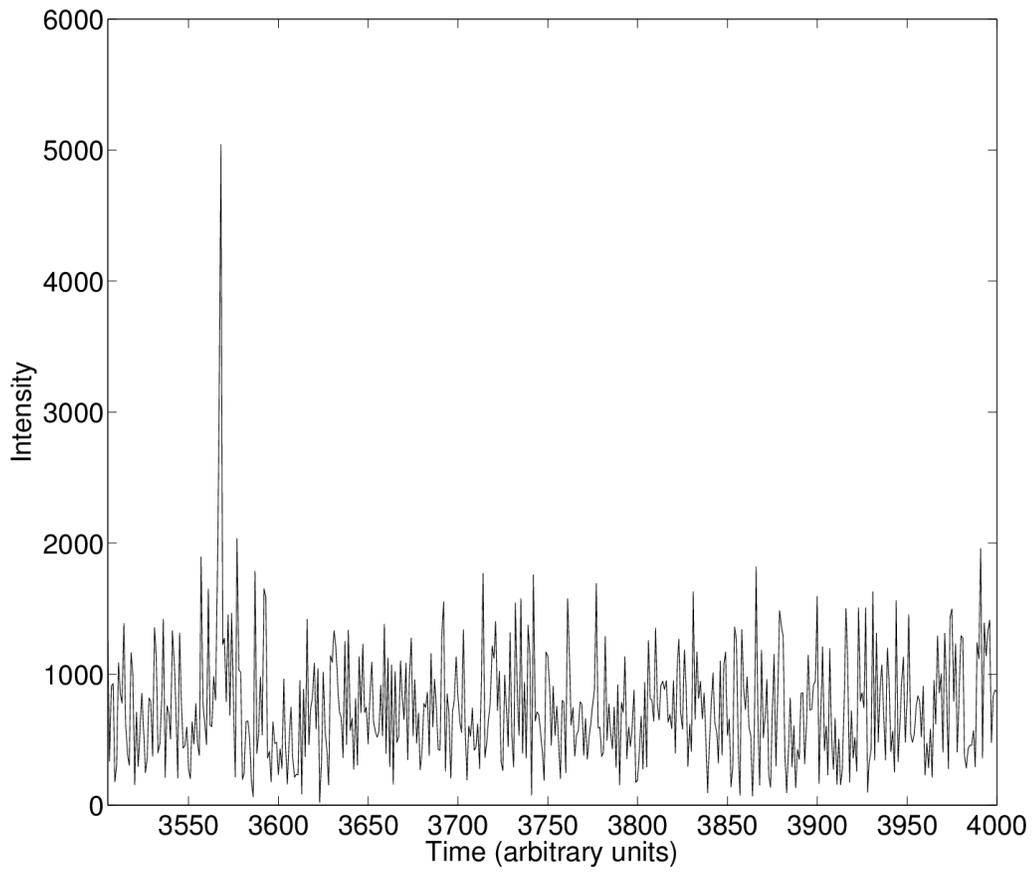

Figure 6

It shows the Fourier transform of the frequency spectrum (Figure 1). The signature of the periodic signal (a spike) is now clearly visible above the noise.